\documentclass[12pt,preprint]{emulateapj}

\usepackage{graphicx}
\usepackage{txfonts}
\usepackage{lscape}
\usepackage{graphicx}
\usepackage{epstopdf}
\usepackage{natbib}
\bibpunct{(}{)}{;}{a}{}{,}
\def\r14{$R^{1/4}$}

\def\kmsMpc{km\,s$^{-1}$\,Mpc$^{-1}$}

\def\mueff{\ifmmode{\mu_{\rm e}}\else{$\mu_{\rm e}$}\fi}
\def\zf{\ifmmode{z_{\rm f}}\else{$z_{\rm f}$}\fi}
\def\Ks{\ifmmode{K_{\rm s}}\else{$K_{\rm s}$}\fi}

\renewcommand{\deg}{\ensuremath{{^\circ}}}

\newcommand{\SExtractor}{\textsc{SExtractor}}
\newcommand{\Hyperz}{\textsc{HyperZ}}

\newcommand{\RS}{\ensuremath{{R_\mathrm{S}}}}
\newcommand{\Pnull}{\ensuremath{{P_\mathrm{null}}}}

\slugcomment{ }

\shorttitle{Nature of the growing bulges at $z<1.3$}
\shortauthors{Dom\'\i nguez-Palmero \& Balcells}

\begin{document}

\title{The nature of growing bulges within $z<1.3$ galaxy disks in the GOODS-N field}

\author{Lilian Dom\'\i nguez-Palmero and 
          \and
          Marc Balcells}
   \affil{Instituto de Astrof\'\i sica de Canarias, 
              E-38200 La Laguna, Tenerife, Spain\\
              \email{ldp@iac.es, balcells@iac.es}}

\begin{abstract}
We analyze central surface brightness $\mu_0$, nuclear and global colors of intermediate redshift disk galaxies. On an apparent-diameter limited sample of 398 galaxies from ACS/HST {\it Great Observatories Origins Deep Survey North} (GOODS-N), we find 131 galaxies with bulges and 214 without. 
Up to $z\sim0.8$, blue, star-forming nuclei are found in galaxies with low $\mu_0$ only; all high-$\mu_0$ nuclei show red, passive colors, so that nuclear and global $(U-B)$ colors strongly correlate with central surface brightness, as found in the local Universe. At $0.8<z<1.3$, a fraction $\sim$27\% of the high-surface brightness nuclei show blue colors, and positive nuclear color gradients.
The associated nuclear star formation must lead to bulge growth inside disks. 
Population modeling suggests that such blue bulges evolve into local pseudobulges rather than classical bulges. We do not find evidence for rejuvenation of classical bulges at the sampled $z$. High luminosity AGNs become common at $0.8<z<1.3$, perhaps pointing to a role of AGN in the growth or star formation truncation of bulges.  
\end{abstract}


\section{Introduction}

In the local Universe, three types of bulges have been defined \citep[see][for recent reviews]{Kormendy04,Athanassoula05}. {\it Classical bulges} are dynamically hot systems with red colors, high concentration, and formation processes perhaps similar to elliptical galaxies. {\it Pseudobulges} \citep{Kormendy93} have lower central densities and bluer colors than classical bulges, often with ongoing star formation. They are believed to form through secular processes that drive gas and stars from the disk inward.  {\it Boxy-peanut shaped bulges}, sometimes considered as a subclass of pseudobulges, are formed via the natural evolution of barred galaxies.

The ideas on the formation histories of bulges can be confronted with direct observations of high-redshift bulge precursors.  
Of critical importance is the search for of strong star formation episodes in the nuclei of galaxies which would qualify as late bulge formation. 
Recent works provide differing accounts of the abundance of blue bulges.  
\citet{Menanteau01} argue for a $30\%-50\%$ of young field spheroids in the two {\it Hubble Deep Fields} (HDF). In contrast, \citet{Koo05} found that $85\%$ of luminous field bulges at $z\sim0.8$ in the Groth strip are nearly as red as local bulges, arguing for very little growth of massive bulges at $z<1$. 
\citet{MacArthur07}, working with GOODS-N field disk galaxies, noted that bulge color is related to mass: the most massive bulges are as red as massive spheroids, perhaps from mass assembly at high redshift; smaller bulges show more diverse star formation histories, with significant star formation at $z<1$. In our own analysis of nuclear colors of disk galaxies at $0<z<1$ in the Groth strip \citep[DB08]{Dominguez08II}, we found that $60\%$ of $0<z<1$ bulges define a passive evolution red sequence (RS); the remaining $40\%$ have bluer colors that may trace ongoing bulge formation/growth.  

For such late-forming bulges two questions arise.  First, what type of $z=0$ bulge (classical, pseudobulge) may be the descendant of an observed star-forming bulge at intermediate $z$. Second, who are their precursors. These may be high-$z$ RS classical bulges rejuvenated due to e.g., gas infall and ensuing star formation. Rejuvenation models have often been advocated \citep[e.g.][]{Ellis01,Koo05}. Alternatively, blue bulges may be the result of centrally concentrated star formation in the disk, perhaps through redistribution of gas through secular or environmental processes.

Distinguishing rejuvenated bulges from bona-fide young bulges growing from their host disks can be done by measuring the relationship between central surface brightness and color. For the general galaxy population in the local Universe, color is strongly correlated with surface density, in the sense that redder galaxies are denser \citep{Kauffmann03dens,Bell00II}, suggesting that galaxies with high surface densities formed the bulk of their stars at earlier epochs than galaxies with lower surface densities. For galaxies with bulges, \citet{Drory07} conclude that galaxies with classical bulges have red global colors and high central surface brightness, whereas galaxies harbouring pseudobulges show bluer global colors and moderate central surface brightness.

In this letter we present the redshift evolution of the relationship between bulge colors and central surface brightness, $\mu_0$. We work with a diameter-limited sample in the redshift range $0.1<z<1.3$. We quantify the $z$-evolution of the $\mu_0$ distribution for bulge and non-bulge galaxies, and investigate the ability of color-$\mu_0$ diagram to segregate different types of bulges at $z\sim1$. Our main result is that, beyond $z\sim0.5-0.8$, blue nuclei appear at increasingly high $\mu_0$, 
tracing episodes of nuclear star formation that lead to bulge growth, most probably by secular processes. Their densities make them candidate precursors of today's pseudobulges, we do not find evidences of rejuvenation of clasical bulges. Interestingly, the AGN fraction increases as well among bulge galaxies at $z>0.8$.

A cosmology with  $\Omega_M=0.3$, $\Omega_\Lambda=0.7$, $H_0=70$ \kmsMpc\ is assumed througout. Magnitudes are expressed in the Vega system.

\begin{table}
\begin{center}
\caption{Bulge sample characteristics}  \label{tab:medians}
\begin{tabular}{l c c c c}
\hline
\hline
 & quantity & $<\mu_0>$ & $\sigma_{\mu_0}$ & $\%$ RS \\
 & (1) & (2) & (3) & (4) \\
\hline
$0.0<z<0.5$ & 50 & 18.79 & 1.12 & 43.6 \\
$0.5<z<0.8$ & 41 & 18.08 & 1.07 & 45.9 \\
$0.8<z<1.3$ & 22 & 17.50 & 0.87 & 31.8 \\
\hline
\hline
\end{tabular}
\end{center}
\begin{footnotetext}
TNote.- (1) Number of galaxies with bulge and without AGN at each redshift range; (2) and (3) median observed $F435W$-band $\mu_0$ and standard deviation; (4) percentage of bulges that belong to the RS, using the cut $(U-B)>0.26$ for bulge galaxies with $\mu_0>19.5$.
\end{footnotetext}
\end{table}

\begin{table}
\begin{center}
\caption{Non-bulge sample characteristics}  \label{tab:mediansNB}
\begin{tabular}{l c c c}
\hline
\hline
 & quantity & $<\mu_0>$ & $\sigma_{\mu_0}$ \\
 & (1) & (2) & (3) \\
\hline
$0.0<z<0.5$ & 92  & 20.57 & 0.99 \\
$0.5<z<0.8$ & 61 & 20.09 & 0.75 \\
$0.8<z<1.3$ & 51  & 19.39 & 0.82 \\
\hline
\hline
\end{tabular}
\end{center}
\begin{footnotetext}
TNote.- (1) Number of galaxies without bulge and without AGN at each redshift range; (2) and (3) median observed $F435W$-band $\mu_0$ and standard deviation.
\end{footnotetext}
\end{table}

\section{Data}
\label{sec:data}

Sample selection followed identical procedures as in DB08, and we refer the reader to \citet{Dominguez08I} for a detailed description, and to \citet{Dominguez08III} for details specific of the present sample. In brief, samples were selected from the ACS/HST GOODS-N \citep{Giavalisco04}, using catalog and image versions v1.1 and v1.0, respectively. The images cover 160 arcmin$^2$ at 0.03 arcsec/pixel through multi-drizzle.  
We started from a diameter-limited parent sample with \SExtractor\ radius above 1.4\arcsec\ (398 galaxies). All have spectroscopic redshifts \citep{Cowie04,Wirth04}, in the range $0.1<z<1.3$. We selected a subsample of galaxies with prominent bulges using the standard {\it mark the disk} method of assuming an exponential profile for the disk. We define a bulge prominence index $\eta$ that measures the excess central surface brightness over the inward extrapolation of the exponential profile of the disk; we take as galaxies with measurable bulges those with $\eta>1$ mag\,arcsec$^{-2}$. The subsample of galaxies with measurable bulges comprises 131 galaxies, after excluding 53 strongly-distorted merger candidates. The remainder 214 galaxies, classified as disk galaxies without bulge, constitute a comparison sample. Galaxies in the latter sample may contain small bulges, unresolved by our images. Indeed, our bulge sample comprises physically large bulges, though not restricted to classical bulges; e.g., simulations show that the prototype peanut-shaped bulge of NGC~5965 would be clearly distinguished out to $z\sim1$ in our survey. See DB08, \S3, for a detailed analysis of selection effects in our bulge sample.  
To investigate selection biases in our diameter-limited sample, we have additionally selected samples by physical radius (above 10.8 kpc); we will show that their behavior is very similar to the angular-limited ones.

The central surface brightness, $\mu_0$ in rest-frame $F435W$, was measured on a central aperture with a physical diameter of 1~kpc, which corresponds to the size of the image PSF for the most distant galaxy in the sample. We tested deconvolution, and bulge-disk decomposition, but both provided dubious results. We calculated $\mu_0$ using the nearest observed band to the rest-frame $F435W$ one at each redshift; it was corrected of the term $(1+z)^4$ of the cosmological dimming, and of inclination and dust, the latter following \citet{Tuffs04} and \citet{Moellenhoff06}.

Nuclear colors were extracted from the same 1~kpc central aperture where we measured $\mu_0$. Integrated galaxy colors were obtained using \textsc{MAG\_BEST} \SExtractor\ magnitudes. For galaxies harbouring a broad-line or a narrow-line AGN, nuclear colors were measured just in the origin of the color profile, and global colors excluded an inner region of radius 2~kpc, to avoid AGN contamination. Rest-frame $(U-B)$ colors were measured using the nearest observed colors at each redshift.

The corrections for filter differences were calculated using SEDs obtained from the best-fit solution, delivered by \Hyperz\ \citep{hyperz}, to photometric points in the bands $F435W,F606W,F775W,F850LP$.

\section{Results and discussion}
\label{sec:results}	

\begin{figure*}
\begin{center}
\includegraphics[angle=0,width=0.95\textwidth]{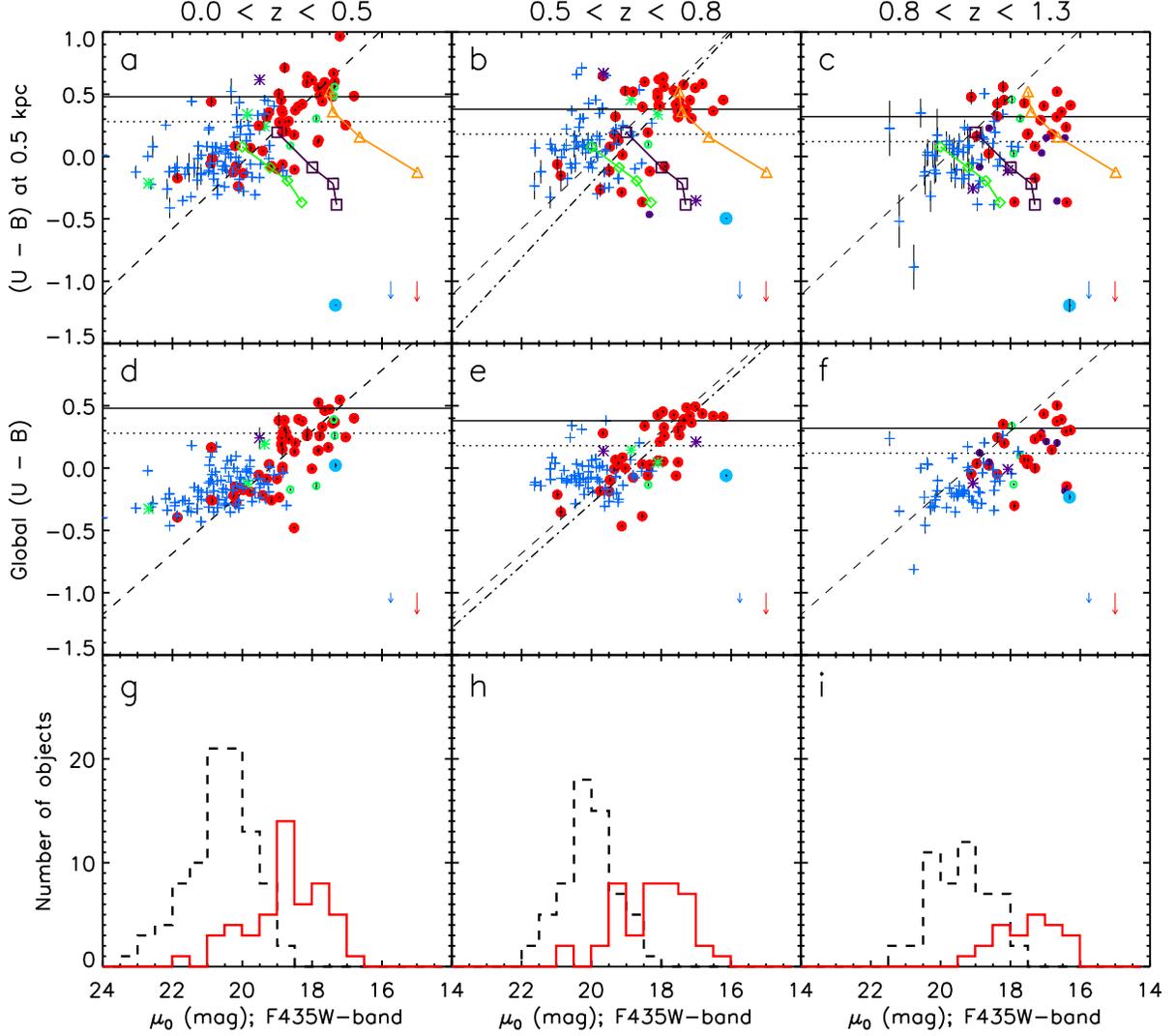}%
\end{center}
\caption{{\bf a,b,c)} and {\bf d,e,f)} galaxy nuclear and global rest-frame $(U-B)$, respectively, vs $F435W$-band $\mu_0$. {\it Intermediate red filled circles}: normal bulge galaxies. {\it Small green open circles}: bulge galaxies hosting a LL AGN. {\it Small dark blue filled circles}: bulge galaxies harbouring a NL AGN. {\it Large blue filled circles}: bulge galaxies hosting a BL AGN. {\it Blue crosses}: normal non-bulge galaxies. {\it Green asterisks}: non-bulge galaxies harbouring a LL AGN. {\it Blue asterisks}: non-bulge galaxies hosting a NL AGN. Colors and $\mu_0$ errors are derived from observed errors. {\it Horizontal solid and dotted lines}: mean color and blue boundary of the RS. {\it Dashed lines}: linear least-squares fits to the color-$\mu_0$ distribution of bulge galaxies in the $z<0.5$ bin. {\it Dot-dashed lines}: least-squares fits at $0.5<z<0.8$ bin. {\it Orange triangle tracks}: CSP model matching RS bulges at $z\sim0$ ($\mu_0=17.5$, $U-B\sim0.5$) with a SFR timescale $\tau=0.8$ Gyr and $\zf=2$ Gyr; each triangle represents the color-$\mu_0$ at redshifts, from right to left, $z=1.2,0.8,0.5,0.1$. {\it Violet square tracks}: CSP model matching $z=0$ bulges with $\mu_0=19$ and $(U-B)\sim0.2$; the model has a SFR $\tau=2$ Gyr and $\zf=1.5$ Gyr. Each square represents the color-$\mu_0$ at redshifts $z=1.2,0.8,0.5,0.1$. {\it Green diamond tracks}: CSP model matching $z=0$ bulges with $\mu_0=20$ and $(U-B)\sim0.1$; the model has a SFR $\tau=2$ Gyr and $\zf=1$ Gyr. Each diamond represents the color-$\mu_0$ at redshifts $z=0.8,0.5,0.3,0.1$. {\it Red vectors}: reddening for bulge galaxies with $B/T=0.4$, central $B$-band opacity $\tau=4$, and inclinations of 40\deg, for nuclear colors ({\bf a,b,c}) and global colors ({\bf d,e,f}). {\it Blue vectors}: reddening for pure disk galaxies with central $B$-band opacity $\tau=4$, and inclinations of of 40\deg, for nuclear and global colors. {\bf g,h,i)} $F435W$-band $\mu_0$ histograms for the three redshifts ranges. {\it Solid red lines}: normal bulge galaxies. {\it Dashed black lines}: normal non-bulge galaxies.}
\label{fig:magsupcenaper}
\end{figure*}

In Fig.~\ref{fig:magsupcenaper}, first and second rows, we plot galaxy nuclear and global rest-frame $(U-B)$ colors vs $\mu_0$, for three redshift ranges: $0.0<z<0.5$, $0.5<z<0.8$ and $0.8<z<1.3$. The horizontal solid and dotted lines trace the mean color of the RS and its blue boundary, respectively, at each redshift bin, from \citet{Kriek08}. The dashed lines trace linear least-squares fits to the color-$\mu_0$ distribution of bulge galaxies in the $z < 0.5$ bin ($U-B=-0.27\mu_0+5.27$ for nuclear colors; $U-B=-0.24\mu_0+4.71$ for global colors); these lines are also plotted in the corresponding higher $z$ bins for reference. The dot-dashed lines trace fits to the bulge galaxy colors in the $0.5<z<0.8$ range ($U-B=-0.30\mu_0+5.81$ for nuclear colors; $U-B=-0.25\mu_0+4.71$ for global colors). 

In the range $0.0<z<0.5$, a strong correlation exists between the $(U-B)$ color and $\mu_0$; both nuclear and global colors are redder in galaxies with higher $\mu_0$. Galaxies without bulges are confined to the blue, faint region of the color-$\mu_0$ diagram, whereas those with bulges cover the entire width of the distribution, with a concentration at the red, high-$\mu_0$ end.  
The trend is very strong (Spearman Rank correlation coefficient $\RS=-0.68$, $\Pnull=2\times10^{-4}\%$ for nuclear colors of bulge galaxies; $\RS=-0.74$, $\Pnull=2\times10^{-5}\%$ for global colors) with one single deviant point, a galaxy harbouring a broad-line AGN. Trends are similarly strong ($\RS=-0.69$, $\Pnull=0.19\%$ for nuclear colors of bulge galaxies) when the sample is limited by physical diameter (\S~\ref{sec:data}). This shows that the color-$\mu_0$ correlation is not driven by large galaxies at the bright end, and small ones at the faint end.

Because red stellar populations, either from old age or from dust, are dimmer than younger populations, the color-$\mu_0$ trend implies that redder bulges have higher central stellar surface mass densities. Hence, at $z<0.5$, central surface density correlates with color, and, hence, with the star formation history of the bulge: denser bulges stopped forming stars in earlier epochs than lower density bulges. 
Not only nuclear colors, but also global colors correlate with central surface density (Fig.~\ref{fig:magsupcenaper}d), as found in the local Universe \citep {Kauffmann03dens}. 
Interestingly, $\mu_0$ for galaxies \textsl{without} bulges also correlate with nuclear colors ($\RS=-0.26$, $\Pnull=1.6\%$) and with global colors ($\RS=-0.37$, $\Pnull=0.06\%$), i.e., the scaling of central surface brightness and color is not entirely due to the presence of the bulge.  

If the two extremes of the color-$\mu_0$ distribution trace classical and pseudobulges \citep{Drory07}, then Fig.~\ref{fig:magsupcenaper}a indicates that the two types of bulges are found up to $z\sim0.5$. Bluer bulges occupy the color-$\mu_0$ locus of star-forming galaxies, and must be forming stars. Notably, all of such galaxies, presenting lower $\mu_0$ than their redder counterparts, must have lower surface densities too. It is unlikely that such blue bulges are rejuvenated classical bulges. If they were, the high density of the underlying classical bulge, plus the high brightness of the burst, would lead to higher surface brightness than red, classical bulges, contrary to what is observed.  We quantify these statements below.  

At the intermediate redshift bin $0.5<z<0.8$ (Fig.~\ref{fig:magsupcenaper}b,e), despite a brighter $\mu_0$ cutoff due to cosmological dimming, the color-$\mu_0$ relation remains in place ($\RS=-0.42[-0.39]$, $\Pnull=0.85\%[2.6\%]$ for nuclear colors of bulge galaxies [selected by physical radius], and $\RS=-0.73$, $\Pnull=3.6\times10^{-4}\%$ for global colors). Linear fits to the nuclear and to the global color-$\mu_0$ distributions yield very similar results to those at $z<0.5$. However, blue deviant nuclei are found around $\mu_0\sim18.5$ and $(U-B)\sim-0.2$ (Fig.~\ref{fig:magsupcenaper}b). The color profiles of those objects (not shown) shows mild blueing toward the center, which provides evidence for enhanced star formation in the nuclear region of the galaxy at redshifts $0.5<z<0.8$.  Two of the four blue deviant bulges may be minor merger remnants, and one hosts a bar, which suggests that these bulges might be growing by secular processes caused by disk instabilities or by satellite accretion \citep{eliche06}. Like in the lower redshift bin, we do not find evidence for rejuvenation of red, classical bulges at $0.5<z<0.8$: the three blue, high-surface brightness nuclei in Figure~\ref{fig:magsupcenaper}b are AGN hosts.

\begin{figure*}
\begin{center}
\begin{minipage}[c]{0.2\textwidth}%
\centering
\includegraphics[angle=0,height=\textwidth]{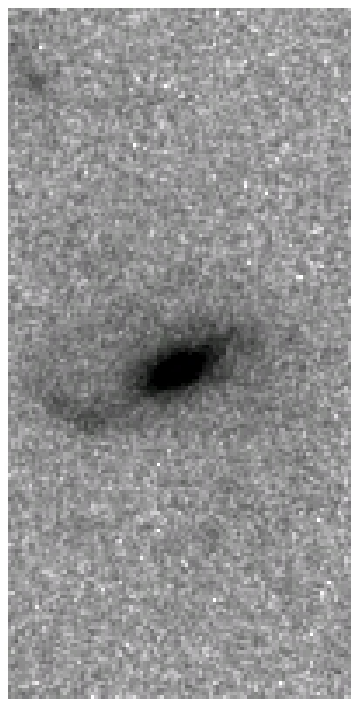}%
\end{minipage}%
\begin{minipage}[c]{0.3\textwidth}%
\includegraphics[angle=0,width=\textwidth]{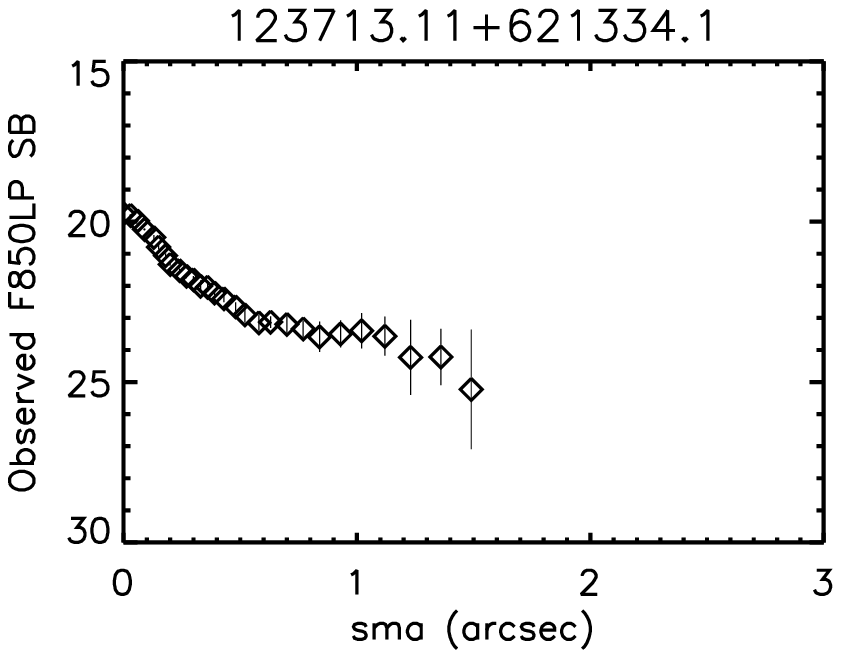}%
\end{minipage}%
\begin{minipage}[c]{0.3\textwidth}%
\includegraphics[angle=0,width=\textwidth]{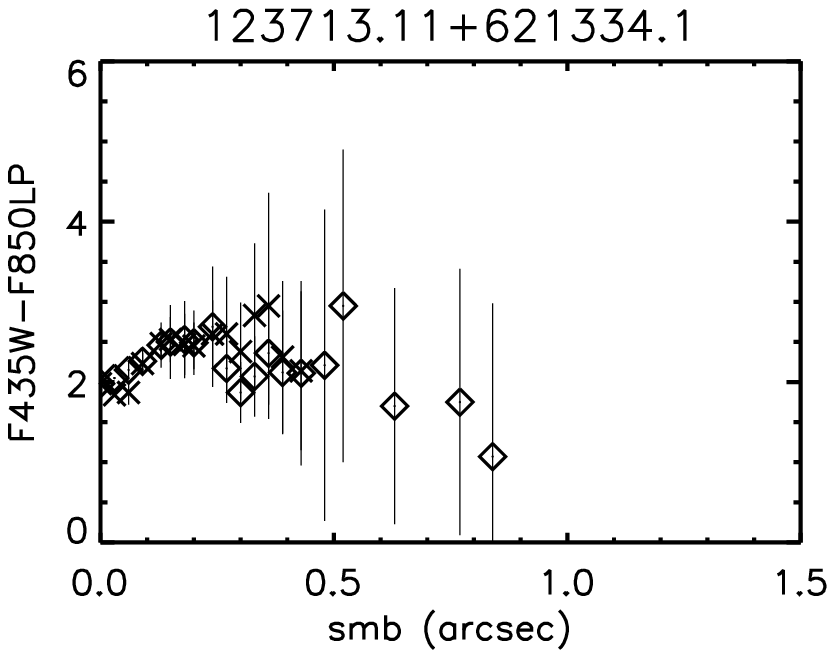}%
\end{minipage}%
\vfill
\begin{minipage}[c]{0.2\textwidth}%
\includegraphics[angle=0,width=\textwidth]{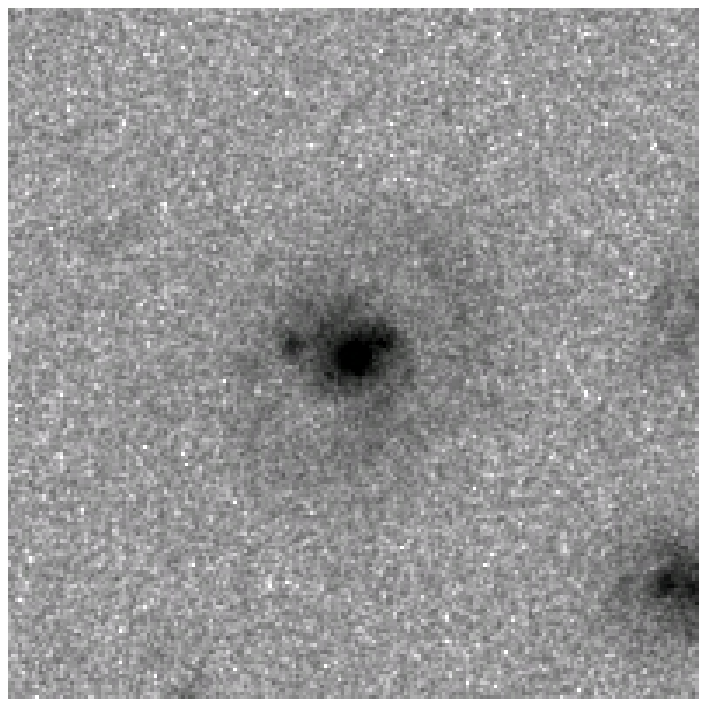}%
\end{minipage}%
\begin{minipage}[c]{0.3\textwidth}%
\includegraphics[angle=0,width=\textwidth]{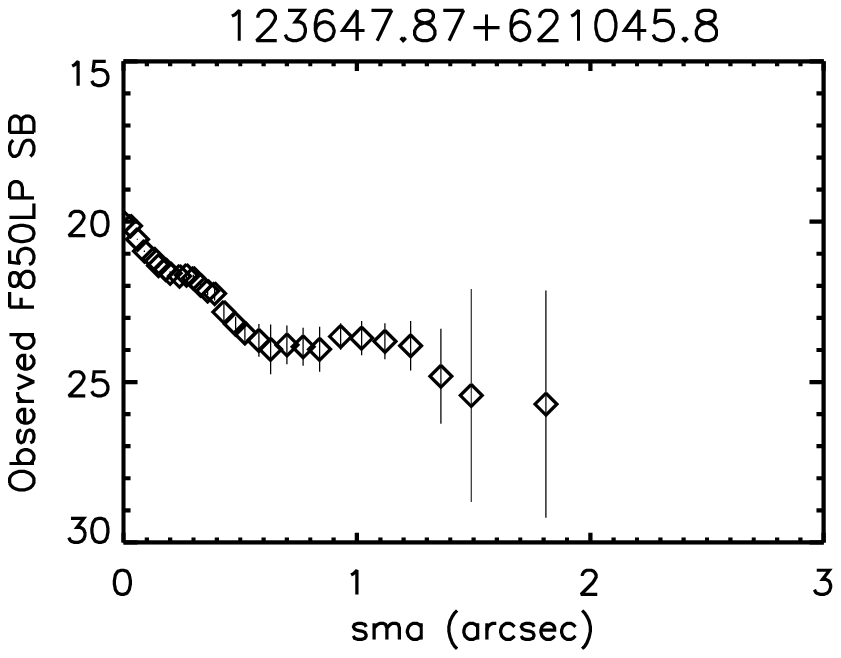}%
\end{minipage}%
\begin{minipage}[c]{0.3\textwidth}%
\includegraphics[angle=0,width=\textwidth]{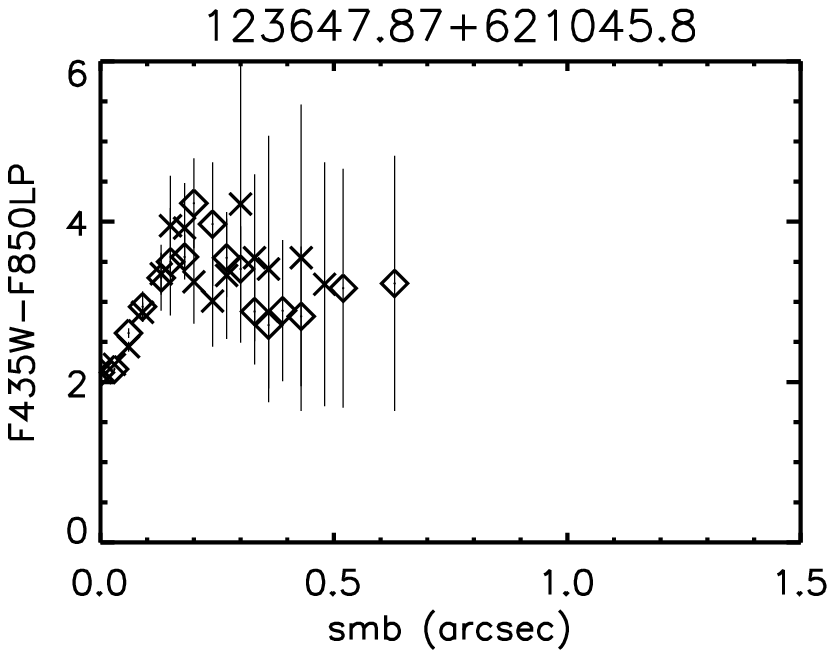}%
\end{minipage}%
\vfill
\begin{minipage}[c]{0.2\textwidth}%
\includegraphics[angle=0,width=\textwidth]{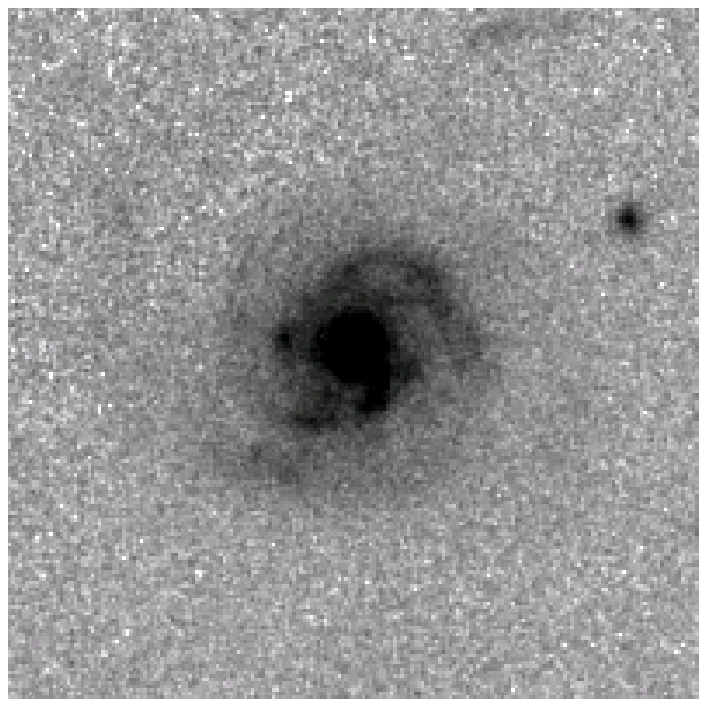}%
\end{minipage}%
\begin{minipage}[c]{0.3\textwidth}%
\includegraphics[angle=0,width=\textwidth]{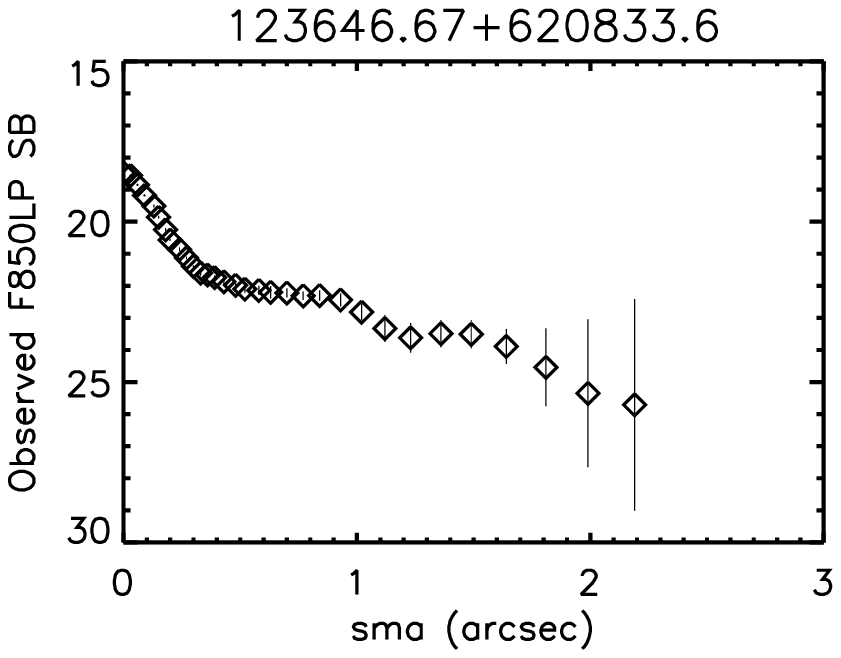}%
\end{minipage}%
\begin{minipage}[c]{0.3\textwidth}%
\includegraphics[angle=0,width=\textwidth]{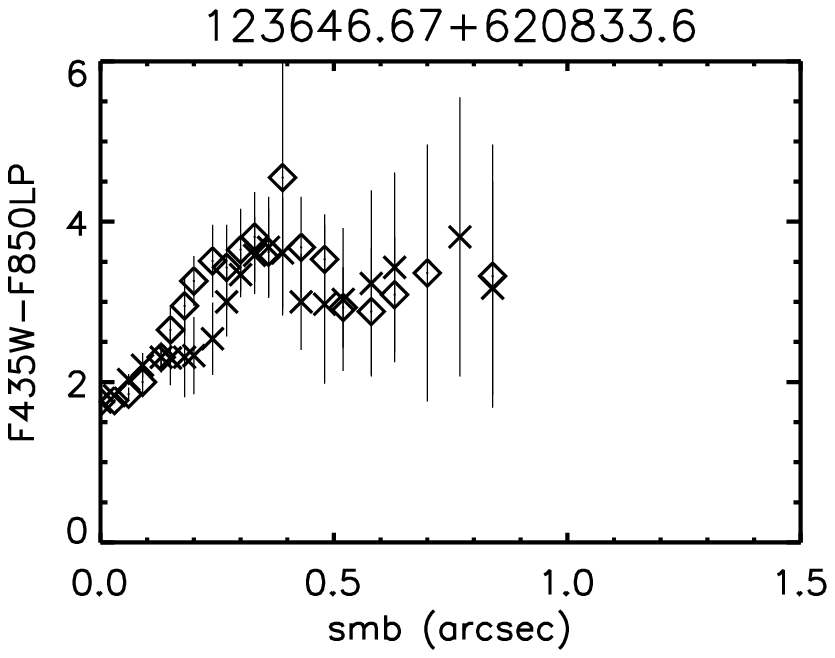}%
\end{minipage}%
\end{center}
\caption{$F850LP$ image, observed $F850LP$ surface brightness profile averaged over both semi-major axes, and observed $(F435W-F850LP)$ color profile on both semi-minor axes for the three non-AGN galaxies at $0.8<z<1.3$ with high $\mu_0$ and nuclear positive color gradient.}
\label{fig:morphology}
\end{figure*}

Finally, at $0.8<z<1.3$, while global colors still correlate with $\mu_0$ (Fig.~\ref{fig:magsupcenaper}f: $\RS=-0.48$, $\Pnull=2.7\%$), nuclear colors (Fig.~\ref{fig:magsupcenaper}c) show a null trend ($\RS=0.10$, $\Pnull=64.3\%$). Figure~\ref{fig:magsupcenaper}c shows a population of blue, high surface brightness nuclei with $\mu_0=16-18$, i.e., as high as the red, classical bulges. In the previous paragraph we already described blue, deviant nuclei at $0.5<z<0.8$, but of a lower $\mu_0=18.0-19.5$. The blue colors and high surface brightness must trace strong episodes of star formation concentrated in the galaxy central regions. The phenomenon may be related to the HDF-N field spheroidals with blue cores \citep{Menanteau01} or to the luminous compact blue galaxies in the HDF-FF \citep{Guzman97}. One of the blue nuclei harbours a narrow-line AGN; the others are undoubtedly stellar. Examples (non-AGN cases) are shown in Figure~\ref{fig:morphology}. One shows signs of interaction; the others are normal spiral galaxies. The color profiles reveal strongly inverted color profiles, bluer inward, in the inner $1-2$ kpc.

Selection effects also contribute against a color-$\mu_0$ correlation in $0.8<z<1.3$. Missing small galaxies, due to our apparent diameter selection, is probably unimportant: the trends and correlation coefficients in Figures~\ref{fig:magsupcenaper}a-c are similar when the samples are selected by physical diameter $D>21.6$ kpc. But galaxies with small \textsl{bulges} may have been classified in the bulge-less class due to resolution and dimming effects (DB08, \S3). Hence, it is plausible that the color-$\mu_0$ correlation survives at $0.8<z<1.3$ if small bulges do exist at those redshifts. In any case, the $0.8<z<1.3$ color-$\mu_0$ distribution remains peculiar for the abundance of blue nuclei as bright as the classical bulges: $27\%$ of the $16<\mu_0<18$ nuclei have colors below $(U-B)=0$.

Figure~\ref{fig:magsupcenaper} clarifies the redshift evolution of the color-$\mu_0$ distribution.  
The linear fits to the bulges, plotted in the diagrams, show that there is no significant change between the ranges $z<0.5$ and $0.5<z<0.8$. But for $0.8<z<1.3$, the distribution lies well below the fit to the first range, with brighter $\mu_0$ and bluer colors. The variation in surface brightness is shown in the $\mu_0$ histograms (Fig.~\ref{fig:magsupcenaper}ghi): the peaks of the distributions of both bulge and non-bulge galaxy samples shift toward brigther $\mu_0$, by $\sim1.2$ mag, from $z\sim0$ to $z\sim1.3$ (see Tables~\ref{tab:medians} and~\ref{tab:mediansNB}; median values do not change significantly if we 
impose a common selection limit at all $z$, $\mu_0=19.5[21]$ for the bulge [non-bulge] sample): Both galaxies with and without bulges were brighter and bluer at $z\sim1$ and faded and reddened, toward $z\sim0$.

Most likely, the centrally-concentrated star forming events contribute to bulge formation inside disks. Who are the precursors, and who are the descendants of such blue, high-surface brightness nuclei? In particular, are they rejuvenated classical bulges that will evolve back into the RS once the burst has faded? We estimate the evolution of starbursting nuclei in color-$\mu_0$ space using composite stellar population (CSP) models from \citet[BC03]{Bruzual03}, with exponentially-decaying star formation rates (SFR), without dust, and build tracks intended to match the span of color-$\mu_0$ at $z\sim0$. For the reddest track, we look for models with $(U-B)\sim0.5$, and assign them a surface brightness of $\mu_0=17.5$. Color and surface brightness are tracked back in time using BC03 and the variation with $z$ of the angular size of the galaxy. For a canonical $\zf=2.0$, and a moderate $\tau=0.8$ Gyr, the model evolves along the orange lines in Fig.~\ref{fig:magsupcenaper}abc, becoming both bluer and brighter as we move back in time.  At $0.8<z<1.3$, the evolution fails to reach the region of the blue, high-surface brightness colors of the starbursting nuclei seen at those redshifts. The bluer colors at $z\sim0.8$ require a younger age, but a younger age would yield too high a surface brightness. Indeed, evolution is not vertical but diagonal in the color-$\mu_0$ plane, as population ageing implies both reddening and dimming.
Taking into account that starbursting nuclei are likely composite old plus young populations does not help, as the underlying, more stable old population would contribute to the surface brightness at all ages, making the problem worse. Hence, the track shows that the blue, high-surface brightness nuclei seen at $0.8<z<1.3$ are not descendants, nor precursors, of the red, high-surface brightness classical bulges.  

For the bluest track, we impose a $z=0$ color of $(U-B)=0.1$, and assign a $z=0$ surface brightness $\mu_0=20$. This track (green line on Fig.~\ref{fig:magsupcenaper}abc) was modeled using a $\zf=1.0$ and a $\tau=2$ Gyr. Blue deviant nuclei with intermediate surface brightness at $0.5<z<0.8$ fit the track evolution. An intermediate model with a slightly brighter $\mu_0$ at $z\sim0$, $\mu_0=19$, and a $\zf=1.5$ (violet line on Fig.~\ref{fig:magsupcenaper}abc), nicely fits the blue, high-surface brightness nuclei at $0.8<z<1.3$. This model links the blue, high-$\mu_0$ deviant bulges at $0.8<z<1.3$ with moderate color, intermediate-$\mu_0$ bulges at $z=0$; it supports the idea that these bulges are precursors of local Universe pseudobulges, rather than old, dense bulges.  

Classifying as AGN those galaxies with $L_X>10^{42}$~erg\,s$^{-1}$ in the 2--8 keV band of \textit{Chandra} \citep{Alexander03,Treister06}, $17\%(5\%)$ of the galaxies with (without) bulges host an AGN in the $0.5<z<1.3$ range.  This highlights that the preference of AGN to live in galaxies with bulges extends up to $z\sim1.3$. In Fig.~\ref{fig:magsupcenaper} we denote with specific symbols three AGN types, namely broad-line (BL), narrow-line (NL), and low-luminosity (LL; no AGN spectral lines), from \citet{Treister06}.  

Except for the three BL AGN, the nuclear colors and $\mu_0$ of galaxies with AGN are in the general region occupied by the non-AGN galaxies -- both in the main color-$\mu_0$ sequence, and in the blue, high-$\mu_0$ region of the starbursting nuclei. Their nuclear colors are likely to be affected by the AGN, especially for the more X-luminous (nuclear colors for our AGN galaxies get bluer with increasing X-ray luminosity). However, the size of the blue nuclear region, and the comparison with synthetic color profiles built with combinations of a disk, a bulge, and a nuclear source with varying colors and luminosity ratios, do suggest that most blue regions are spatially resolved, hence that star formation coexists with the AGN.  

The trigger of the AGN activity in these galaxies is unclear. 
Morphologies for the three BL AGN hosts reveal a normal, bulge-dominated spiral structure, with a central bar in two of them. The NL AGN galaxies show a more mixed variety of types, including early-type galaxies, normal and barred spirals, and distorted galaxies.  Hence, AGN activity is not associated with merging or interacting situations, as noted by  \citet{Grogin05}.  

\section{Conclusions}
\label{sec:conclusions}

In the GOODS-N field, $27\%$ of very blue, high-$\mu_0$ bulges in $0.8<z<1.3$ trace an epoch of important bulge formation inside disks. The fraction of blue cores is close to that for HDF field spheroidals \citep{Menanteau01}. Similar but weaker nuclear star formation is present at $0.5<z<0.8$, pointing at bulge-forming events that become weaker with cosmic time, and lead to a mass-age relation for bulges \citep{MacArthur07}. The data imply important bulge growth at $z<1$. Models predict that such starbursting nuclei evolve into moderate surface-brightness, intermediate color $z=0$ pseudobulges rather than classical bulges, and argue against rejuvenation processes for $z\sim1$ dense and old bulges. The latter are already abundant at $0.8<z<1.3$, as found by \citep{Koo05}.

At lower redshifts, $0.0<z<0.8$, the correlation of nuclear and global colors with $\mu_0$, and with central density, suggests that classical and pseudobulges exist at least up to $z=0.8$.

The high frecuency of AGNs in $z>0.5$ galaxies with bulges suggests a link between AGN activity and bulge growth in morphologically normal disk galaxies at $z\sim1$.

\acknowledgments
We thank the anonymous referee for suggestions that improved the paper and Ignacio Trujillo and Mercedes Prieto for useful discussions. This work is based on archival data from the HST, obtained from the data archive at STScI, which is operated by AURA under NASA contract NAS5-26555. Support has been provided by the Spanish Programa Nacional de Astronom\'\i a y Astrof\'\i sica through project number AYA2006-12955 and through the Consolider-Ingenio 2010 Program grant CSD2006-00070: First Science with the GTC (http://www.iac.es/consolider-ingenio-gtc/).

\clearpage

\end{document}